\documentclass{aa}
\usepackage{graphicx}

\begin{document}

\title{Photometric parameters of edge-on galaxies from 2MASS observations}

\author{D.Bizyaev\inst{1} \and S.Mitronova\inst{2} }

\offprints{D.Bizyaev}

\institute{Sternberg Astronomical Institute, Universitetsky prospect 13,
119899, Moscow, Russia;\\ 
Isaac Newton Institute of Chile, Moscow Branch\\
\email{dmbiz@sai.msu.ru}
 \and
Special Astrophysical Observatory of RAS, pos. Nizhnij Arkhyz,357147, 
Karachaevo-Cherkessia, Russia\\
       \email{mit@sao.ru}
    }

\date{Received 04 January 2002; accepted 23 April 2002}

\abstract{

To analyze the vertical structure of edge-on galaxies, we have used images
of a large uniform sample of flat galaxies that have been taken during the
2MASS all-sky survey. The photometric parameters, such as the radial scale
length, the vertical scale height, and the deprojected central surface
brightness of galactic disks have been obtained. We find a strong
correlation between the central surface brightness and the ratio of the
vertical scale height to the vertical scale length: the thinner the galaxy,
the lower the central surface brightness of its disk. The vertical scale 
height does not increase systematically with the distance from the galaxy
center in the frames of this sample.

   \keywords{ galaxies: structure -- galaxies: photometry}
}

\maketitle

\section{Introduction}

The study of edge-on galaxies provides a unique possibility to obtain
information about vertical structure in galactic disks. Beginning with the
papers of van der Kruit \& Searle (\cite{KS81a}, \cite{KS81b},
\cite{KS82}) the investigation of edge-on galaxies continues
today in optical bands (see van der Kruit \cite{K01} and references therein)
as well as at radio wavelengths (Matthews et al, \cite{MGvD99}, van der
Kruit et al, \cite{KJKF01}). These studies help us to understand the laws
governing the distribution of stellar and gaseous components of disks, and
shed light on the role of dark halos in the evolution of spiral galaxies.

The main difficulty in studying edge-on disks in the optical is the need to
take into account internal extinction by dust inside the disks. Extinction 
values can be enormous in the plane of a galaxy (of the order of a few tens
of magnitudes for our Galaxy) and even far outside the plane they are
substantial (see Xilouris et al, \cite{Xetc99}). This is why it is   
preferable to use red and infrared data to investigate the structure of
edge-on galaxies.

The 2MASS (Two Micron All Sky Survey) image library provides a sample of
edge-on galaxies with near-infrared photometric data.  Unfortunately the
exposure time for 2MASS objects was too small to image the external parts of
spiral galaxies, but it allows us to study the structures of their thin
disks.

In this paper we use 2MASS data to obtain information about photometric
parameters of stellar disks: their radial scale length, vertical scale
height and deprojected central surface brightness for as many galaxies as
possible. We choose the infrared photometric band $K_s$ for this study as it
is the 2MASS band least influenced by dust extinction.

\section{Sample of galaxies}

The data from the 2MASS Public Release Image Server were used (see Nikolaev
et al., \cite{N_2000} for details of the images).  All images and
calibration data were collected using the web-based interface provided by
the NASA/IPAC Extragalactic Database (http://irsa.ipac.caltech.edu/
applications/2MASS/ReleaseVis/). Image tiles containing objects from the
Revised Flat Galaxy Catalog (Karachentsev et al., \cite{RFGC}, hereafter
RFGC) were chosen for this analysis. This gives some guarantee that the
galaxies chosen all have thin and highly-inclined disks.

Our initial sample included more than 700 objects from the RFGC catalog
which had already been detected in 2MASS survey. An examination of 
these images shows that there are only 153 objects whose major axes, $A$,
span more than 40\arcsec (in $K_s$ filter). We note that this value is
related to the visible size of objects shown in the 2MASS frames. Such a
widely used value as the isophotal diameter $D_{25}$ is generally about
three times greater than the maximum size of the same objects in $K_s$
frames. Flat galaxies that have a value of $A$ close to 40\arcsec are
marginally acceptable for the present analysis. The most critical parameter
for the vertical structure analysis is the object size in the vertical
direction i.e. parallel to the minor axis.  Galaxies with $A < 40\arcsec$ do
not allow us to analyze clearly the vertical cuts.

Finally, we refined this sample to a subsample consisting of 60 flat
galaxies whose major axis size is more than 1\arcmin. Galaxies in this
subsample seem to be more reliable for our analysis due to bigger angular diameter 
and will be examined together with the main sample of 153 galaxies to look for a
possible difference in photometric parameters between the main sample and
the subsample. This allows us to estimate possible systematic errors caused
by the small size of some objects.

\section{A method for obtaining the radial scale length,
the vertical scale height and the central surface brightness of stellar disks }

We follow the classic way to investigate the vertical structure of
edge-on disks in spiral galaxies (van der Kruit \& Searle, \cite{KS81a},
\cite{KS81b}, \cite{KS82}, de Grijs \& van der Kruit \cite{deGK96}). It
allows us to trace the behavior of the scale height at different distances
from the center of the galaxy analyzing each photometric cut separately. In
previous studies roughly ten cuts were made parallel to the minor axis of 
each galaxy, and a few parallel to its major axis.

In this paper we analyze a few tens of cuts spaced at equal intervals
parallel to the minor axis of the disk. To estimate the radial scale length
and the central surface brightness we made two cuts parallel to the major
axis located at a certain distance from the galactic plane, as in the plane
the value of internal extinction remains relatively large even in the IR
spectral range.

A preliminary inspection of each frame was made to make an initial guess at
the position of galactic center, as well as at the approximate size and
orientation of the ellipse outlining the visible size of a galaxy. The
ellipse is drawn along the faintest isophote seen in the image.  The cuts
were done parallel to the major and the minor axes of the ellipse.  The cuts
parallel to the major axis were gradually shifted our away from the major
axis (in both directions) by 12-15\% of the minor axis size in order to
examine the surface brightness profile along the major axis.
The number of cuts parallel to the minor axis was chosen to be around 20-30,
thus covering most of the disk. The 2MASS survey is not sensitive to the
very faint parts of galaxies, so our results are related mostly to the thin
stellar disks. Hence, emission from the thick stellar disk
might be neglected near the galactic plane.

The equations used to fit the parameters of each cut assume an exponential law
for the distribution of luminosity volume density in the radial direction
$\rho_L(r)$ and an isothermal law for the distribution in the vertical
direction. In this case a general form of surface brightness distribution
$I$ on the plane of a sky $(X,Y)$ is

\begin{equation}
\label{e1}
I(X, Y) = \int \rho_L (r,z) dl ~~~,
\end{equation}

\noindent where 
\begin{equation}
\label{e2}
\rho_L (r,z) = \rho_{L0} \, \exp \left( -\frac{|r|}{R_e} \right)
\mathrm{sech}^2 \left( \frac{|z|}{z_0} \right) ~~~,
\end{equation}

\noindent $r$ is the radial distance from the center of a galaxy, $z$ is the
distance from the galactic plane, $z_0$ and $R_e$ denote the vertical scale height and
the radial scale length respectively. Integration of (\ref{e1}) proceeds along the
line of sight.  One can find the deprojected central surface brightness
$S_0$ by integrating (\ref{e2}) along the vertical axis.

In the first-order approximation we assume $z_0$ to be independent of radius
$r$, as has been previously noted by many authors (see e.g. van der Kruit
\& Searle, \cite{KS81a}, \cite{KS81b}, de Grijs \& Peletier,
\cite{dGP97}). To take into account the radial variability of the vertical
scale height, we must know a priori the form of $z_0(r)$. More generally, we
can assume that $z_0$ is a monotonous function of $r$. Then, using our
approximation we can determine whether $z_0$ varies significantly with
radial distances or not.

We use Eq. (\ref{e1}) by fixing the Y coordinate  and drawing two cuts
parallel to the major axis to obtain the values of $R_e$ and
$S_0$. In a similar way, we fit the cuts drawn parallel to the minor axis to find the values
of $z_0$. In this case we add a variable "shift term" $\delta Y$ to find the
declination of the center of each cut from the galactic plane.

In this study we do not consider the possibility that some disks may not be
90\degr inclined, because we find that most objects from the RFGC catalog
have inclinations that are close to 90\degr. A more detailed study that
includes the non-90\degr inclined disks is problematic, since we usually do
not see a dust layer on most of the 2MASS galactic images taken in the
$K_s$-band.

While analyzing the radial surface brightness profiles, we exclude their
central parts (typically 1/3 - 1/4 of the maximal extent of the radial
profile) to decrease the influence of the bulge light on our results.

The smearing effect caused by the Earth's atmosphere affects the values of
the measured scales by increasing the value of the scale height. To take this
effect into account, we convolved the fitting functions with a gaussian. The
mean FWHM value of the PSF was about 3\arcsec during the survey so we have
chosen 3\arcsec as the FWHM of the gaussian. Finally the functions were
fitted to the extracted profiles by the least-squares method.

The flux calibration of the central deprojected surface brightnesses $S_0$
is available from equations given by Nikolaev et al. (\cite{N_2000}). 
Finally, we corrected the $S_0$ values for the Galactic extinction
(using the data from LEDA, the Lyon-Meudon extragalactic database).

We analyzed the set of cuts parallel to the major and minor axes for each
galaxy in our 2MASS sample and obtained the values of $S_0$, $R_e$,
$z_0(r)$, and $\delta Y$ as the output. The value of $Y_0$ was used to
correct the position angle of the ellipse outlining the faintest isophote in
the image of a galaxy, if it was needed, and to correct
the final value of $z_0$ for each distance from the center $r$.  Finally, we
take the value of $z_0$ as the median value of all scale heights and the
value of $R_e$ as the average value of all scale lengths for each galaxy.
The averaging of scale length/height allow us to
avoid the influence of stars and bad pixels projected onto the images on the
final values of $R_e$ and $z_0$.

The output photometric parameters are presented in Table 1. It contains (1)
the name of the galaxy according to RFGC catalog, (2) the distance in Mpc
adopted in our paper, (3) the radial scale length in kpc, (4) the central
surface brightness reduced to the face-on inclination in mag/$\sq \arcsec$,
(5) the vertical scale height in kpc, and (6) the label "x" marking the
galaxy that belongs to the subsample. 

\section{Results and Discussion}

\begin{figure}
   \centering
   \includegraphics[width=8.5cm]{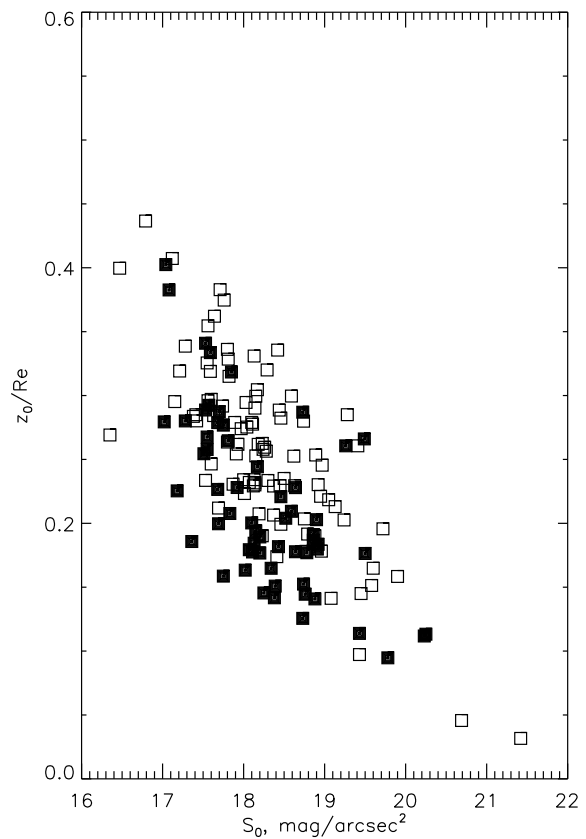}
      \caption{
Relation between the corrected central surface brightness, $S_0$, and the
ratio of the vertical scale height to the radial scale length $z_0 / R_e$ for the sample of 153
galaxies. The subsample of galaxies is denoted by the filled
squares.
              }
         \label{F1}
   \end{figure}

Fig.~\ref{F1} shows the relation between the corrected central surface
brightness $S_0$ and the ratio $z_0 / R_e$ 
of the scale height to the scale length of a disk.
Open squares in Fig.~\ref{F1} denote our main sample of galaxies. Filled
squares show the subsample of galaxies. The linear fit gives
the equation $z_0 / R_e = 1.30 - 0.059 \cdot S_0$ for the 
subsample of 60 galaxies in Fig.~\ref{F1}.

As it is seen in  Fig.~\ref{F1}, there is a strong correlation between $S_0$
and $z_0 / R_e$. Thicker disks have higher values of central surface
brightness, and vice versa. The scatter in the points for the
the main sample is larger than the scatter for the subsample. This indicates
a typical uncertainty in $z_0/R_e$ and $S_0$ due to the low quality of some
images. A better correlation between these values for the more resolved
galaxies gives us hope for even better results when the higher resolution
photometric data will be available.

The systematic difference between galaxies of different sizes (open versus
filled squares) is well seen in Fig.~\ref{F1}. It reflects the
overestimation of $z_0$ for the faintest galaxies of the main sample.

As was noted by Gerritsen \& de Blok (\cite{GdB99}), the low surface
brightness galaxies (LSB) must be relatively thinner than the normal (HSB)
galaxies. The most natural explanation of this feature is that a dark halo
does contribute a substantial part of the mass in LSB galaxies (de Blok \&
McGaugh, \cite{dBmG97}). A shallower central part of the rotation curve is a
typical feature of LSB galaxies, which requires a large fraction of dark
matter (van den Bosch et al., \cite{BDdB00}). Fig.~\ref{F1} shows that the
values of deprojected central surface brightness of disks span about 2.5
magnitudes (for the subsample of galaxies). This agrees well
with the observational data published by de Jong (\cite{dJ96a}), Tully \&
Verheijen (\cite{TV97}) for almost face-on disks.  These authors
show that the typical difference in $S_0$ values between LSB and HSB
galaxies is of the order of 2 magnitudes.

From the $K_s$-band central surface brightness of 18.2 mag/$\sq \arcsec$ and
the $B$-band central surface brightness of 21.65 mag/$\sq \arcsec$ (Freeman,
\cite{F70}) we infer the $(B-K_s) \approx 3.5$ for the central parts of
deprojected face-on disks. This is in good agreement with the values of
$S_0$ in Fig.~\ref{F1}. Fainter disks in Fig.~\ref{F1} have $S_0$
values that differ from the typical Freeman value in $K_s$-band by 1.5 - 2
magnitudes. These faint disks appear to be the LSB disks in our sample.

   \begin{figure}
   \centering
   \includegraphics[width=8.5cm]{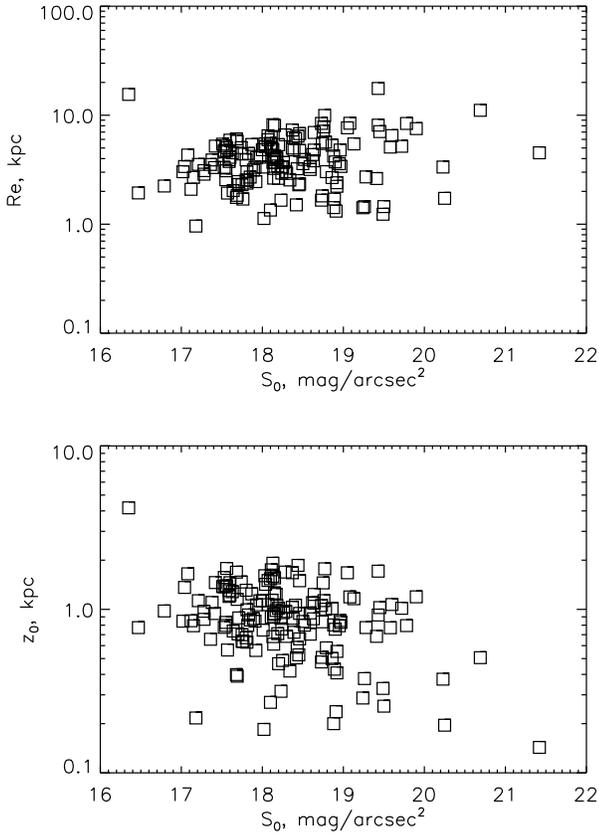}
      \caption{
The linear values of radial scale length $R_e$ (upper figure) and vertical
scale height $z_0$ (lower figure) for galaxies of different central surface
brightnesses $S_0$.
              }
         \label{F2}
   \end{figure} 

Using galactic distances based on radial velocities reduced for Galactic
Standard of Rest movement ($V_{GSR}$ taken from LEDA), we compare the
linear values of $z_0$ and $R_e$ for galaxies with different central surface brightnesses 
(see Fig.~\ref{F2}). We use a Hubble constant of 75 $\mathrm{km}~ \mathrm{s}^{-1}
\mathrm{Mpc}^{-1}$ throughout this paper. As can be seen in Fig.~\ref{F2},
neither the linear value of $R_e$ nor that of $z_0$ tends to show a
systematic difference between LSB and HSB disks. Note that the errors in
distances lead to additional scatter of the points in Fig.~\ref{F2},
especially for nearby galaxies. A more sophisticated analysis of
distance-related values based on new distance estimates is needed. This
question will be addressed in the next paper.

Nevertheless, the correlations between $S_0$ and each of scales $z_0$ and
$R_e$ (Fig.~\ref{F2}) are much worse than the correlation between $S_0$ and
the ratio of scales $z_0/R_e$ (Fig.~\ref{F1}). It implies that the latter correlation 
is based on a more significant physical foundation
than the correlations between $S_0$ and each of scales $z_0$ and
$R_e$. A pronounced correlation between $S_0$ and
$z_0/R_e$ implies the importance of dark halos in galaxies with
low values of $S_0$.  The gravity of the halo does not let the disk grow
thicker in this case.

   \begin{figure}   
   \centering                 
   \includegraphics[width=8.5cm]{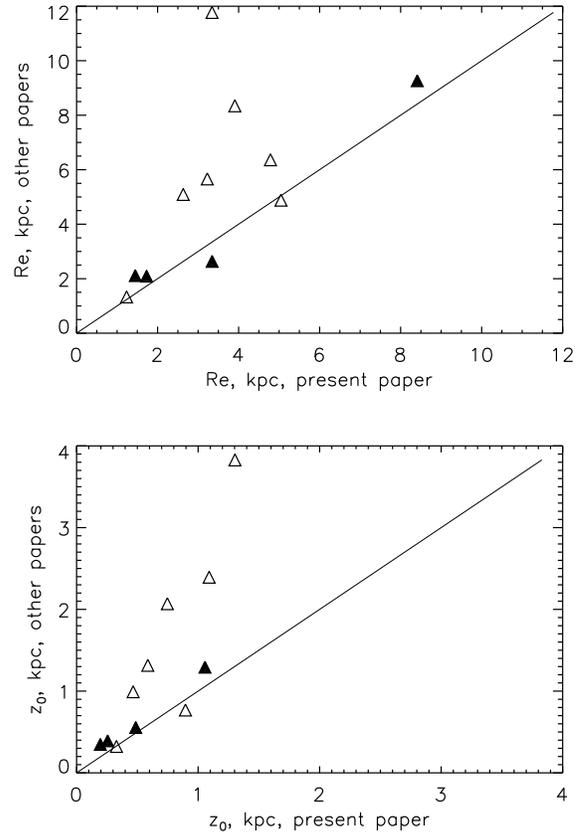} 
      \caption{
The comparison of the radial scale lengths (upper frame) and the vertical
scale heights (lower frame) between the published data and the data
presented in this paper.  Filled triangles denote the scales that were
obtained using K' data. Open triangles are related mostly to R-band
observations. Solid lines show the case of the equality of the scales.
              }
         \label{F3}
   \end{figure}    

Eleven galaxies from our sample have the scales measured and published by
Barteldrees \& Dettmar (\cite{BD94}) and Schwarzkopf \& Dettmar
(\cite{Sw2000_1}). Four of them were observed in the K'-band, which is very
close to $K_s$ where our estimates have been done. Fig.~\ref{F3} shows a
comparison of radial scale lengths (upper frame) and vertical scale heights
(lower frame) made for these cases. Distance-dependent values for galaxies
from other papers are corrected to our distance scale. Filled triangles
denote the scales that have been obtained using the K'-band data. Open
triangles are related mostly to the R-band observations. As Fig.~\ref{F3}
shows, scale lengths and scale heights are as a rule larger in the R-band
and pretty close to our values in the K'-band. One of the galaxies, UGC
7321, was investigated by Matthews (\cite{M00}) in the K-band. She found
that the scale length and scale height are equal to 2.0 and 0.19 kpc
respectively (here the scale height of the thin galactic disk is cited). Our
values are 1.73 and 0.20 kpc, which do not differ significantly. Two other
galaxies NGC 4244 and NGC 5907 were investigated by van der Kruit \& Searle
(\cite{KS82}) using photographic plates in the J-band (close to Johnson's
B). Using distances to galaxies from the present paper (3.6 and 10.9 Mpc
respectively) we can compare our scale lengths/scale heights. The published
values are 1.87/0.41 kpc and 5.65/0.82 kpc after reducing to our distance
scale, whereas our values are 1.23/0.33 kpc and 3.35/0.49 kpc respectively.
The values are of the same order and a systematic difference is expected
because of usually shorter scales in near-infrared filter bands. Finally,
the scales for another galaxy, UGC 11194, were found using I-band
observations and published by Bizyaev (\cite{Biz2000}). The scales are
4.48/0.98 kpc in contrast to 5.04/0.90 kpc found in this paper. The scales
for the same galaxy were published by Reshetnikov \& Combes (\cite{RC97}).
After correcting to our distance scale they are 4.0/1.15 kpc.

   \begin{figure}   
   \centering                 
   \includegraphics[width=8.5cm]{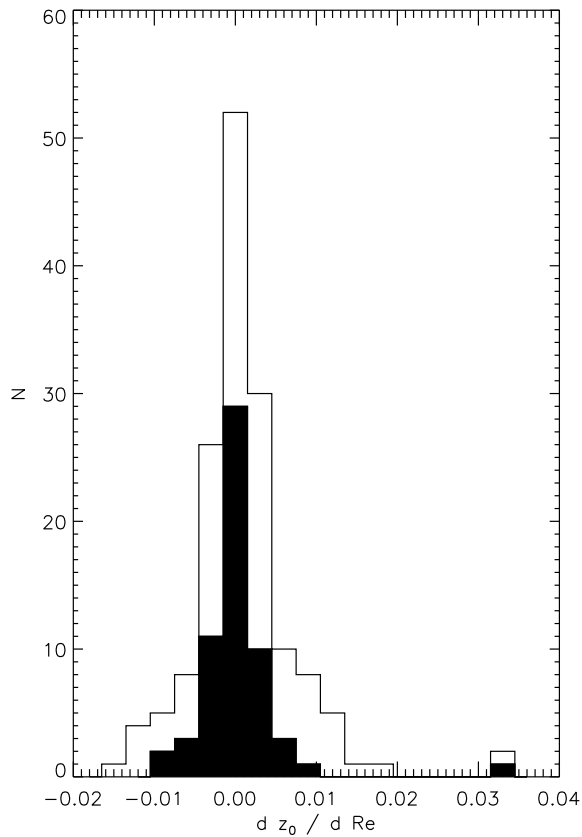} 
      \caption{
The radial gradient of  the vertical
scale height $d z_0 / d r$ for our sample of galaxies. The average value of
$d z_0 / d r$ is very close to zero. The standard deviation of $d
z_0 / d r$ is equal to 0.0063. The subsample of galaxies is   
highlighted by the filled part of histogram.
              }
         \label{F4}
   \end{figure}    

In spite of the low accuracy achieved for each individual cut, we are able
to estimate roughly the radial trend of $z_0(r)$. To do so, we exclude a few
values of scale height that deviate from the general trend by more than 3
$\sigma$.  We fit the values of $z_0$ to a linear function of r.  The
histogram in Fig.~\ref{F4} shows the results of the fitting. The gradient $d
z_0 / d r$ is presented in dimensionless values. The subsample of galaxies
is shown by the filled part of the histogram in Fig.~\ref{F4}.

As one can see in Fig.~\ref{F4}, the average value of $d z_0 / d r$ is very
close to zero (0.001), which is in agreement with the conclusions of many
previous studies (see previously cited references). The subsample of
galaxies has an average value of $d z_0 / d r$= 0.0003 (i.e. zero). The
standard deviation of $d z_0 / d r$ in Fig.~\ref{F4} is equal to 0.006 for
the whole sample and 0.005 for the 60 largest galaxies in this sample. It
gives a 10\% change of the measured scale height on distances of 3.5 $R_e$
from the center, i.e. at the very edge of a stellar disk adopting a typical
value of $z_0 / R_e$ from Fig.~\ref{F1}. Note that the real change of $z_0$
along the radius needed to produce the 10\% trend may be 2-3 times more
significant due to projection of different parts of the disk on the line of
sight.

\section{Conclusions}

We analyzed the vertical and the radial distributions of near-infrared
surface brightness in disks of flat galaxies observed during the 2MASS
survey. A strong dependence of the ratio $z_0/R_e$ of the vertical scale
heights to the radial scale lengths on the deprojected central surface
brightness $S_0$ is inferred.  Galaxies with lower central surface
brightnesses look thinner and have lower values of $z_0/R_e$. The linear
value of the radial scale length appear to be independent of the central
surface brightness, the same conclusion is true for the vertical scale
height.  We can also conclude that the vertical scale height of thin stellar
galactic disks is almost independent of radius for galaxies in our sample
and has almost the same value over a wide range of distances from the
center.

\begin{acknowledgements}

The authors would like to thank I.D. Karachentsev and A.V.Zasov for fruitful
discussions. D.B. thanks A.Moiseev and D.Makarov for their hospitality at
Special Astrophysical Observatory RAS and discussions. We thank Verne Smith
for comments on this paper. We'd like to thank Eduard Vorobyov for his help in
text improvement. We also thank the referee whose remarks improved the paper
content.

This work was partially supported by the Russian Foundation for Basic
Research via grant 01-02-17597.

We made use of data products from the Two Micron All Sky Survey, which is a
joint project of the University of Massachusetts and the Infrared Processing
and Analysis Center/California Institute of Technology, funded by the
National Aeronautics and Space Administration and the National Science
Foundation. NED is operated by the Jet Propulsion Laboratory, California
Institute of Technology, under contract with the National Aeronautics and
Space Administration. We also have made use of the LEDA database
(http://leda.univ-lyon1.fr).

\end{acknowledgements}


\newpage
\begin{center}
Table 1. Photometric parameters of disks of galaxies.

\begin{tabular}{rrrcrr}
\hline
Name  &     D     & $R_e$   & $S_0$    &  $z_0$ & $Note^1$\\
RFGC  &     Mpc   &   kpc   &  mag/$\sq \arcsec$ & kpc &      \\
\hline
  95  &    13.10  &   1.45  &  19.50   &   0.26 & x    \\
 139  &    72.34  &   3.18  &  18.59   &   0.95 &      \\
 176  &    72.01  &   5.66  &  18.78   &   1.00 & x    \\
 183  &    69.80  &   8.11  &  19.43   &   0.92 & x    \\
 206  &    62.16  &   1.94  &  16.47   &   0.77 &      \\
 282  &    91.56  &   3.69  &  18.15   &   0.93 &      \\
 355  &    74.42  &   4.79  &  18.96   &   0.85 &      \\
 363  &    75.23  &   5.36  &  18.20   &   1.02 & x    \\
 420  &    74.57  &   3.10  &  17.28   &   0.87 & x    \\
 444  &   122.18  &   6.10  &  18.41   &   1.06 &      \\
 485  &   110.64  &   5.45  &  19.13   &   1.16 &      \\
 504  &    50.76  &   4.80  &  18.39   &   0.72 & x    \\
 517  &    71.07  &   5.19  &  19.72   &   1.02 &      \\
 507  &    53.84  &   2.38  &  17.81   &   0.63 & x    \\
 538  &    26.71  &   1.66  &  18.73   &   0.48 & x    \\
 544  &    84.44  &   3.33  &  17.91   &   0.85 &      \\
 551  &   148.94  &   5.02  &  17.56   &   1.78 &      \\
 561  &   145.61  &   4.70  &  17.56   &   1.39 &      \\
 586  &    56.42  &   2.04  &  17.64   &   0.74 &      \\
 603  &    70.77  &   2.24  &  16.79   &   0.98 &      \\
 609  &    84.09  &   4.60  &  17.53   &   1.57 & x    \\
 642  &    34.56  &   2.55  &  18.34   &   0.42 & x    \\
 653  &   162.75  &  17.61  &  19.43   &   1.71 &      \\
 671  &   143.69  &   8.18  &  18.13   &   1.92 &      \\
 702  &    36.65  &   1.70  &  18.89   &   0.43 &      \\
 722  &    24.11  &   2.23  &  18.92   &   0.41 & x    \\
 744  &   124.66  &   4.96  &  18.12   &   1.14 &      \\
 757  &   131.95  &   5.51  &  18.10   &   1.54 &      \\
 765  &   121.79  &   6.06  &  17.68   &   1.69 & x    \\
 826  &   115.75  &   7.66  &  19.05   &   1.67 &      \\
 882  &    80.19  &   3.09  &  17.55   &   0.80 & x    \\
 895  &    15.23  &   1.45  &  19.26   &   0.38 & x    \\
 902  &    58.95  &   2.71  &  17.85   &   0.86 & x    \\
 914  &   125.78  &   4.03  &  18.17   &   0.98 & x    \\
1047  &    36.33  &   1.82  &  18.74   &   0.51 &      \\
1049  &    13.39  &   0.96  &  17.18   &   0.22 & x    \\
1128  &    51.89  &   4.41  &  17.75   &   0.70 & x    \\
1135  &    63.82  &   2.30  &  18.45   &   0.53 &      \\
1140  &    63.88  &   3.32  &  17.53   &   0.78 &      \\
1143  &    58.62  &   3.73  &  18.90   &   0.76 & x    \\
1159  &    85.70  &   6.78  &  18.46   &   1.50 & x    \\
1167  &    52.15  &   3.37  &  18.59   &   0.71 & x    \\
1194  &    29.44  &   2.30  &  17.75   &   0.64 & x    \\
1206  &    32.57  &   1.84  &  17.69   &   0.39 &      \\
1244  &    41.43  &   4.79  &  18.64   &   1.09 & x    \\
1263  &    63.54  &   3.34  &  18.01   &   0.75 &      \\
1329  &    54.63  &   2.62  &  19.41   &   0.68 &      \\
1339  &    73.42  &   7.83  &  18.76   &   1.13 & x    \\
1349  &    56.14  &   2.69  &  17.15   &   0.79 &      \\
1421  &    19.98  &   1.76  &  17.68   &   0.40 & x    \\
1431  &    58.34  &   5.42  &  17.51   &   1.38 & x    \\
1459  &   102.94  &   3.14  &  17.89   &   0.88 &      \\
\hline
\end{tabular}

\noindent $^1$ - {\small "x" means part of the more reliable subsample (see text).}

\newpage
Table 1, Continue

\begin{tabular}{rrrrrr}
\hline

1499  &    65.25  &   2.09  &  17.12   &   0.85 &      \\
1500  &    22.07  &   3.03  &  17.02   &   0.85 & x    \\
1502  &   119.76  &   4.82  &  18.00   &   1.13 &      \\
1536  &   113.49  &   4.03  &  18.18   &   1.05 &      \\
1548  &    55.37  &   2.34  &  18.46   &   0.66 &      \\
1624  &    60.63  &   3.59  &  18.95   &   0.79 &      \\
1627  &    96.27  &   3.35  &  18.23   &   0.88 &      \\
1670  &    99.85  &   8.41  &  18.73   &   1.06 & x    \\
1691  &    81.79  &   3.82  &  18.63   &   0.88 &      \\
1692  &    93.00  &   4.36  &  18.62   &   1.10 &      \\
1723  &    64.25  &   3.43  &  18.19   &   0.71 &      \\
1754  &   110.32  &   6.60  &  18.28   &   1.69 &      \\
1789  &    27.04  &   1.66  &  18.23   &   0.32 &      \\
1792  &    14.61  &   1.13  &  18.02   &   0.18 & x    \\
1872  &   103.64  &   3.76  &  17.59   &   1.20 &      \\
1901  &    83.97  &   3.92  &  18.14   &   0.90 &      \\
1904  &   129.33  &   7.13  &  18.75   &   1.45 &      \\
1906  &    13.72  &   1.32  &  18.91   &   0.24 & x    \\
1928  &    80.12  &   4.07  &  17.93   &   1.07 &      \\
1932  &    38.21  &   2.40  &  18.92   &   0.55 &      \\
1945  &    14.14  &   1.35  &  18.10   &   0.27 & x    \\
2026  &   101.81  &   4.12  &  17.97   &   1.13 &      \\
2044  &    51.01  &   2.65  &  18.14   &   0.61 &      \\
2068  &    75.97  &   3.39  &  18.97   &   0.83 &      \\
2097  &    68.96  &   5.10  &  19.58   &   0.77 &      \\
2100  &    86.45  &   4.21  &  18.15   &   1.26 &      \\
2162  &    21.70  &   1.42  &  19.24   &   0.29 &      \\
2171  &    47.64  &   2.72  &  19.28   &   0.77 &      \\
2174  &   100.18  &   3.68  &  18.26   &   0.96 &      \\
2239  &    11.40  &   1.42  &  18.88   &   0.20 & x    \\
2245  &     3.61  &   1.23  &  19.49   &   0.33 & x    \\
2246  &    10.00  &   1.73  &  20.25   &   0.20 & x    \\
2257  &    50.64  &   2.46  &  17.70   &   0.71 & x    \\
2296  &   106.92  &   7.53  &  19.90   &   1.19 &      \\
2312  &    83.28  &   3.31  &  17.41   &   0.94 &      \\
2315  &    13.81  &   8.40  &  19.78   &   0.80 & x    \\
2336  &     7.58  &   4.52  &  21.42   &   0.14 &      \\
2373  &    68.45  &   5.21  &  17.55   &   1.39 & x    \\
2376  &    90.16  &   4.06  &  18.17   &   1.24 &      \\
2380  &    47.61  &   3.93  &  18.52   &   0.80 & x    \\
2399  &    28.67  &   3.35  &  20.23   &   0.37 & x    \\
2418  &   117.08  &   8.44  &  19.08   &   1.19 &      \\
2425  &    36.76  &   1.70  &  17.76   &   0.64 &      \\
2473  &   146.73  &   3.89  &  17.80   &   1.31 &      \\
2477  &    95.17  &   4.75  &  17.53   &   1.37 & x    \\
2517  &    35.04  &   2.68  &  18.86   &   0.50 & x    \\
2528  &    35.18  &   5.50  &  18.74   &   0.84 & x    \\
2568  &    79.64  &   2.88  &  17.28   &   0.97 &      \\
2611  &   101.78  &   7.07  &  19.45   &   1.03 &      \\
2679  &    71.90  &   3.59  &  18.50   &   0.84 &      \\
2682  &    33.94  &   3.38  &  17.04   &   1.36 & x    \\
2715  &   136.69  &   6.47  &  18.07   &   1.50 &      \\
2747  &    39.53  &   6.22  &  18.38   &   0.88 & x    \\
2756  &    35.23  &   2.63  &  18.20   &   0.47 & x    \\
2826  &    31.39  &   2.46  &  17.92   &   0.56 & x    \\
2835  &    45.32  &   3.53  &  17.36   &   0.66 & x    \\
\hline
\end{tabular}

\newpage
Table 1, Continue

\begin{tabular}{rrrrrr}
\hline
2860  &    28.59  &   1.93  &  17.57   &   0.56 & x    \\
2945  &   153.26  &   9.95  &  18.77   &   1.77 &      \\
2946  &    10.90  &   3.35  &  18.25   &   0.49 & x    \\
2966  &    92.63  &   4.74  &  18.46   &   0.95 &      \\
2994  &    47.61  &   4.31  &  17.08   &   1.65 & x    \\
3004  &   127.28  &   6.29  &  18.11   &   1.75 &      \\
3006  &    43.30  &   3.04  &  18.79   &   0.58 &      \\
3085  &    62.95  &   2.55  &  17.80   &   0.67 &      \\
3094  &   120.43  &   8.03  &  18.15   &   1.56 & x    \\
3098  &   132.44  &   4.81  &  18.13   &   1.59 &      \\
3106  &   125.47  &   5.45  &  18.03   &   1.61 &      \\
3114  &    32.16  &   1.50  &  18.42   &   0.50 &      \\
3227  &   115.45  &   5.17  &  18.04   &   1.42 &      \\
3240  &   128.98  &   5.04  &  17.74   &   1.47 &      \\
3313  &    56.19  &   2.54  &  17.82   &   0.80 &      \\
3352  &    72.73  &   3.25  &  18.14   &   0.94 &      \\
3378  &    75.42  &   2.94  &  18.24   &   0.76 &      \\
3430  &    59.13  &   4.99  &  18.37   &   1.03 &      \\
3455  &    65.01  &   2.93  &  18.30   &   0.68 &      \\
3477  &   143.56  &   6.44  &  18.44   &   1.86 &      \\
3480  &   150.33  &   7.31  &  18.37   &   1.67 &      \\
3507  &    78.03  &   3.91  &  17.59   &   1.30 & x    \\
3580  &    72.82  &   5.30  &  18.86   &   1.01 & x    \\
3614  &    83.47  &   4.22  &  18.88   &   0.80 &      \\
3658  &   109.38  &   4.51  &  17.63   &   1.28 &      \\
3762  &   117.38  &   5.21  &  17.42   &   1.46 &      \\
3793  &    46.09  &   3.70  &  18.13   &   0.68 & x    \\
3863  &    77.13  &   5.96  &  18.07   &   1.07 & x    \\
3903  &    35.29  &   3.23  &  18.43   &   0.59 & x    \\
3926  &    67.84  &   5.05  &  18.11   &   0.90 & x    \\
3984  &   117.45  &   6.50  &  19.60   &   1.07 &      \\
3985  &    89.81  &   3.03  &  17.81   &   0.99 &      \\
4004  &    99.54  &   6.95  &  18.64   &   1.23 & x    \\
4013  &    43.43  &  11.09  &  20.69   &   0.51 &      \\
4043  &    99.73  &   3.54  &  17.21   &   1.13 &      \\
4051  &    90.88  &   3.03  &  18.29   &   0.97 &      \\
4076  &    97.13  &   4.46  &  17.83   &   0.93 & x    \\
4078  &   118.28  &   5.77  &  17.69   &   1.15 & x    \\
4092  &    93.99  &   2.56  &  17.55   &   0.83 &      \\
4103  &   145.16  &   5.91  &  17.60   &   1.46 &      \\
4106  &    98.73  &   4.11  &  17.60   &   1.22 &      \\
4110  &    82.41  &  15.53  &  16.35   &   4.18 &      \\
4136  &    70.31  &   2.32  &  17.71   &   0.89 &      \\
4165  &    92.36  &   3.89  &  17.38   &   1.10 &      \\
4171  &   101.53  &   5.39  &  17.87   &   1.24 &      \\
\hline
\end{tabular}

\end{center}


\begin{thebibliography}{}

\bibitem[1994]{BD94}
Barteldrees, A., \& Dettmar, R.-J. 1994, A\&AS, 103, 475


\bibitem[2000]{Biz2000}
Bizyaev, D. 2000, Soviet Astron. Lett., 26, 219


\bibitem[1996]{deGK96}
de Grijs, R., \& van der Kruit, P. 1996, A\&AS, 117, 19
 
 
\bibitem[1997]{dGP97}
de Grijs, R., \& Peletier, R. 1997, A\&A, 320, L21

 
\bibitem[1997]{dBmG97}
de Blok, W., \& McGaugh, S. 1997, MNRAS, 290, 533


\bibitem[1996a]{dJ96a}
de Jong, R. 1996, A\&A, 313, 45
 
\bibitem[1996b]{dJ96b}
de Jong, R. 1996, A\&A, 313, 377


\bibitem[1970]{F70}
Freeman, K. 1970, ApJ, 160, 811

\bibitem[1999]{GdB99}
Gerritsen, J., \& de Blok, W. 1999, A\&A, 342, 655

\bibitem[1999]{RFGC}
Karachentsev, I., Karachentseva, V., Kudrya, Y., et al. 1999, Bull. of
Special Astrophys. Obs., 47, 5

\bibitem[1999]{MGvD99}
Matthews, L., Gallagher, J., \& van Driel, W. 1999, AJ, 118, 2751

\bibitem[2000]{M00}
Matthews L., 2000, AJ, 120, 1764

\bibitem[2000]{N_2000}
Nikolaev, S., Weinberg, M., Strutskie, M. et al. 2000, AJ, 120, 3340

 
\bibitem[2000]{PDL00}
Pohlen, M., Dettmar, R.-J., \&  Lutticke, R. 2000, A\&A, 357, L1

\bibitem[1997]{RC97}
Reshetnikov, V., \& Combes., F. 1997, A\&A, 324, 80

\bibitem[2000]{Sw2000_1}
Schwarzkopf, U., \& Dettmar, R.-J. 2000 A\&AS, 144, 85

\bibitem[1997]{TV97}
Tully, R., \& Verheijen, M. 1997, ApJ, 484, 145

\bibitem[2000]{BDdB00}
van den Bosch, F., Robertson, B., Dalcanton, \& J., de Blok, W. 2000, AJ, 119,
1579

\bibitem[1981a]{KS81a} 
van der Kruit, P., \& Searle, L. 1981a, A\&A, 95, 105

\bibitem[1981b]{KS81b}
van der Kruit, P., \& Searle, L. 1981b, A\&A, 95, 116

\bibitem[1982]{KS82}
van der Kruit, P., \& Searle, L. 1982, A\&A, 110, 61


\bibitem[2001]{K01}
van der Kruit, P. 2001, astro-ph/0109480 

\bibitem[2001]{KJKF01}
van der Kruit, P., Jimenez-Vicente, J., Kregel, M., \& Freeman, K. 
2001, A\&A, 379, 374


\bibitem[1999]{Xetc99}
Xilouris, E., Byun, Y., Kylafis, N. et al. 
1999, A\&A, 344, 868


\end{thebibliography}
\end{document}